# TURING PATTERNS IN NONLINEAR OPTICS

**Kestutis Staliunas**

Physikalisch Technische Bundesanstalt, 38116 Braunschweig, Germany

**Víctor J.Sánchez-Morcillo**

Departament D'Òptica, Universitat de València, Dr.Moliner 50, E-46100 Burjassot, Spain

## Abstract

The phenomenon of pattern formation in nonlinear optical resonators is commonly related to an off-resonance excitation mechanism, where patterns occur due to mismatch between the excitation and resonance frequency. In this paper we show that the patterns in nonlinear optics can also occur due to the interplay between diffractions of coupled field components. The reported mechanism is analogous to that of local activation and lateral inhibition found in reaction-diffusion systems by Turing. We study concretely the degenerate optical parametric oscillators. A local activator-lateral inhibitor mechanism is responsible for generation of Turing patterns in form of hexagons.

PACS numbers 42.65.-k, 42-65.Sf, 42.65.Yj

A well-known transverse pattern formation mechanism in broad aperture lasers and other nonlinear resonators is off-resonance excitation. If the central frequency of the gain line of the laser $w_A$ is larger than the resonator resonance frequency $w_R$, then the excess of frequency $\Delta w = w_A - w_R$ can cause transverse (spatial) modulation of the laser fields, with characteristic transverse wavenumber $k_\perp$ obeying a dispersion $ak_\perp^2 = \Delta w$, where $a$ is the diffraction coefficient of the field in the resonator. The patterns occurring in such a way play the role of a ``bridge`` between the external excitation frequency and the internal resonance frequency, and enable maximum energy transfer through the system.

The off-resonance excitation mechanism, which require $\Delta w > 0$, is responsible for occurrence of patterns not only in lasers but also in other nonlinear optical systems, like externally injected nonlinear resonators [1,2], photorefractive oscillators [3], or degenerate and nondegenerate optical parametric oscillators (DOPOs and OPOs) [4,5].

The pattern formation in passive nonlinear systems [6] is also essentially an off-resonance phenomenon. The nonlinear focusing create the off-axis fields components, which are then amplified due to the resonator detuning. The characteristic transverse wavelength of the emerging patterns scales with the square root of diffraction constant $a^{1/2}$ (the transverse wavenumbers participate only in a combination $ak_\perp^2$ in the linear stability analysis of off-resonance patterns).

On the other hand, the pattern formation mechanism discovered by Turing for reaction-diffusion systems [7], has a different origin from the above described. The Turing mechanism is related with the interplay between the diffusions of interacting components. The coupling between a strongly diffusing (lateral) inhibitor and a weakly diffusing (local) activator can cause the pattern formation in systems described by reaction-diffusion equations. This situation occurs in chemical kinetics, as in the Brusselator model [8], and in biological problems, as in the Gierer-Meinhardt model [9]. All these systems share common features, as (*i*) the pattern formation threshold is a simple function of the diffusion coefficients and a control parameter, (*ii*) an intrinsic wavelength of the patterns that is independent of the geometry of the system, and (*iii*) the existence of a minimum value for the ratio between diffusions for the pattern formation process be possible [10].

A simple (linearized) representation of such a reaction-diffusion equations displaying Turing instability is given by the model [11]

$$\P_t u_1 = c_1 u_1 - c_2 u_2 + D_1 \nabla^2 u_1, \qquad (1a)$$
$$\P_t u_2 = c_3 u_1 - c_4 u_2 + D_2 \nabla^2 u_2. \qquad (1b)$$

Here $u_1$ is the activator and $u_2$ is the inhibitor, with corresponding diffusion coefficients $D_1$ and $D_2$ respectively. The particular form of cross coupling matrix ($c_i$ being positive values) leads to maximal amplification of the wavenumbers $k$ obeying $2k^2 = c_1/D_1 - c_4/D_2$, as follows from the stability analysis of Eqs.(1) [11].

Turing pattern formation mechanism has not been clearly identified in nonlinear optics up to now. The main reason is that, in the models previously considered, the diffractions of the interacting fields, that represent the nonlocality in optical systems (as diffusions are in the Turing problem) had a fixed ratio, and then its role in the pattern formation process could not be explored.

In this paper we study the doubly resonant DOPO, and show that not only the conventional off-resonance patterns, but also Turing-like patterns can be excited. In a DOPO the excitation frequency is equal to half of the pump radiation frequency $w_A = w_p/2$, and its mismatch from $w_R$ leads to the same off-resonance pattern formation mechanism as discussed above for lasers. However, in addition to the known off-resonance patterns [12] also Turing patterns occur, due to interplay between diffractions of interacting fields. They are analogous to the Turing patterns in reaction-diffusion occurring due to interplay between diffusions of coupled components.

The mean-field dynamical equations for signal (subharmonics) $A_1(\mathbf{r},t)$ and pump waves $A_0(\mathbf{r},t)$ are [12]

$$\partial_t A_1 = g_1[-(1+i\Delta_1)A_1 + A_1^* A_0 + ia_1 \nabla^2 A_1], \qquad (2a)$$
$$\partial_t A_0 = g_0[-(1+i\Delta_0)A_0 + E - A_1^2 + ia_0 \nabla^2 A_0], \qquad (2b)$$

where $E$ is the amplitude of the (external) pumping field, $\Delta_1$ and $\Delta_0$ are the detunings of the fields with respect to the resonance frequencies of the resonators, $g_1$ and $g_0$ are the decay rates and $a_1$ and $a_0$ are diffraction coefficients of the signal and pump waves in the resonator. Throughout the paper the case of resonant pump $\Delta_0 = 0$ and equal decay rates $g_1 = g_0$ is considered for simplicity.

It is well known that, when the optical cavity is formed by plane mirrors, the diffraction coefficients of signal and pump fields in a DOPO are linked through the relation $a_1 = 2a_0$, as a result of the phase-matching condition [12]. On the other hand, the curvature of the mirrors modifies the structure of the transverse Laplacian, and imposes a particular boundary conditions not present in the model, which has been derived assuming the mean-field approximation. However, the use of resonators with curved mirrors close to a confocal (or more generally, self-imaging) configuration is nearly equivalent to the plane mirror case (with high Fresnel number and high level degeneracy of transverse modes), in which the diffraction coefficient depends on the deviation of the resonator length from confocality. In particular, a exactly confocal resonator is diffractionless (every ray has the same optical length in one round-trip in the resonator). This equivalence has been shown analytically in [13], using a propagation matrix approach, and experimentally in [14], where the observed patterns have been compared with the solutions of a mean-field model, with good agreement.

In the present case, we assume that each field resonates in a near-self-imaging cavity, with different lengths. Then, the diffraction coefficients can take independent values, being their ratio a parameter of the system. This configuration allows to explore the role that diffraction plays on the pattern formation properties in this system, and to show some similarities with reaction-diffusion systems and the Turing mechanism.

An essential difference between reaction-diffusion systems and nonlinear optics is the nonlocality: in the classical reaction-diffusion systems (*e.g.* (1)) the nonlocality is diffusion, while in optics (*e.g.* (2)) the nonlocality is predominantly diffraction. We generalize therefore the Turing pattern formation mechanism of local activation and lateral inhibition (LALI) to the general case of nonlocalities, and we expect that the Turing patterns can occur not only in the diffusive case of model system (1), but also for diffractive DOPO equations (2). We call the occurring patterns by LALI patterns, in order to distinguish them from the off-resonance patterns. The above qualitative arguments suggest that the LALI instability could be observed when the ratio between pump (inhibitor) and subharmonic (activator) diffractions in a DOPO reaches a critical value.

The rest of the paper is organized as follows: we first perform the linear stability analysis of the nontrivial (nonzero) solution of (2) and we show the growth of unstable Turing modes for sufficiently large ratio of pump and subharmonic diffractions $a_0/a_1$. The growing Turing modes give rise to hexagonal patterns as shown by numerical integration of (2). The spatial wavenumber and the threshold conditions for the hexagonal Turing patterns are analytically evaluated, by means of asymptotic expansions. Next, we analyze the case when the pump diffraction is too small to cause instability (the case of under-developed LALI). This leads to a nontrivial spatial distribution of noisy patterns: in the far field the radiation is distributed not on the central spot, as occurs for zero pump diffraction, but forms a ring, as follows from the numerical solutions of the DOPO Langevin equations.

A spatially homogeneous solution $\overline{A}_0, \overline{A}_1$ of Eqs.(2) can be obtained by elimination of the pump field $A_0$ from (2a), and using the ansatz $A_1 = \overline{A}_1 = A\exp(i\varphi)$ for the subharmonics. Then the intensity and the phase of the signal field are given by [15,16]

$$A^2 = -1 + \sqrt{E^2 - \Delta_1^2}, \sin(2\varphi) = -\frac{\Delta_1}{E} \qquad (3)$$

with an additional constraint for the phase: $\cos(2\varphi) > 0$. The stationary pump value corresponding to (2) is $\overline{A}_0 = (E - A_1^2)$.

Below we consider that the spatial coordinates are normalized to $\sqrt{a_1}$. This is equivalent to set $a_1 = 1$ and $a_0 = a$ in Eqs.(2), being $a = a_0/a_1$ the diffraction coefficient.

In order to investigate the stability of the homogeneous solution (3) against space-dependent perturbations we perturb it with a perturbation of the form $d\mathbf{A}(\mathbf{r},t) \propto \exp(\lambda t + i\mathbf{k}\cdot\mathbf{r})$, where $d\mathbf{A} = (dA_0, dA_0^*, dA_1, dA_1^*)$. This stability analysis has been performed previously [16-18], but with the diffraction coefficient restricted to the value $a = 1/2$. We investigate here the instabilities for arbitrary values of the diffraction parameter $a$.

The linear stability analysis leads to a fourth order polynomial in the eigenvalues, and then to explicit, although complicated, mathematical expressions for the growth rates $\lambda(k)$. In fig.1 we represent the real part of λ against the perturbation wavenumber, $k = |\mathbf{k}|$, for three different values of the diffraction and a fixed positive value of signal detuning. The parameters are such, that the off-resonance detuning instability does not occur (signal detuning is positive). For zero pump diffraction $a = 0$ (dotted curve), the homogeneous solution is stable. The off-axis modes are strongly damped, and no LALI instability occurs. For a diffraction value $a = 1/2$ (dashed curve in Fig.1), the homogeneous solution is stable, the off-axis modes $k \neq 0$ are damped, but the damping around some wavenumbers is weak. This corresponds to the situation where LALI instability is sensible, but below the threshold (under-developed LALI). Increasing the value of the diffraction, the largest real part of the eigenvalue grows, until it becomes positive at a critical wavenumber $k = k_c$. This situation is shown by the continuous curve in Fig.1, obtained for a diffraction value $a = 10$.

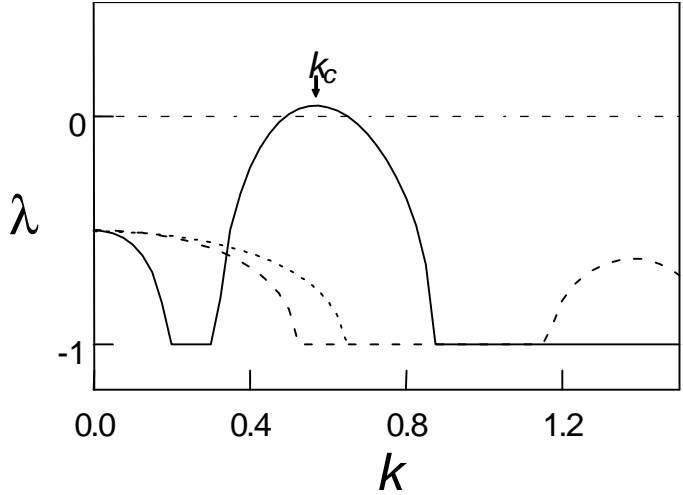

*Fig.1. Real part of the eigenvalues depending on the perturbation wavenumber, for different values of the diffraction; a = 0 (pointed curve), a = 0.5 (dashed curve) and a = 10 (solid curve). The other parameters are $D_1 = 1$, $D_0 = 0$, $E = 2.5$.*

It is easy to show that, for $a = 0$, the only possible instability is the off-resonance one, with a wavenumber given by $k^2 = -\Delta_1$, proving the requirement that pump diffraction, and therefore the nonlocal interplay between fields, is needed for the existence of Turing patterns.

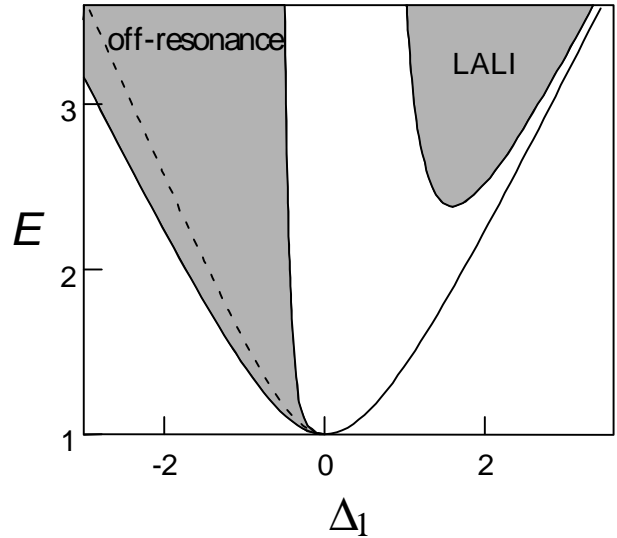

*Fig.2. Instability regions in the parameter space ($D_1,E$) for nonzero diffraction a = 5, evaluated from the linear stability analysis. There are two instability regions: for negative detunings, the traditional off-resonance instability; for positive detuning the LALI instability is present.*

At the threshold of the pattern-forming instability the real part of the eigenvalue of the wavenumber with maximum growth is zero. In Fig.2, the off-resonance and LALI instability regions are plotted in the parameter space $(\Delta_1, E)$ for a given value of diffraction. The regions are well separated in the parameter space, and therefore should be attributed to different mechanisms. The off-resonance instability exists for all values of the pump above threshold, whereas the LALI instability appears only at some critical pump value, that depends on the diffraction *a*. We note that the pump diffraction not only creates the LALI instability, but also modifies the off-resonance instability range, as seen from Fig.2. For zero pump diffraction the off-resonance instability occurs in between the dashed curve, and the left solid curve corresponding to the neutral stability line, as follows from the standard instability analysis. The pump diffraction increases significantly the off-resonance instability region. However, the spatial scale of the off-resonance pattern is not modified by the presence of pump diffraction.

Another important feature revealing the different nature of the patterns in both sides of resonance is the corresponding wavelength. In the case of off-resonance patterns, it mainly depends on the resonator detuning and diffraction coefficient of the signal wave. In contrast, the wavelength of the pattern in the LALI region depends essentially on the pump and on the ratio of the diffraction coefficients $a$, and very weakly on the resonator detuning. This behavior is shown in Fig.3, where the squared wavenumber that maximizes the growth rate is plotted against the detuning (full line). The broken part of the curve, in the neighbourhood of the resonance, corresponds to negative eigenvalues.

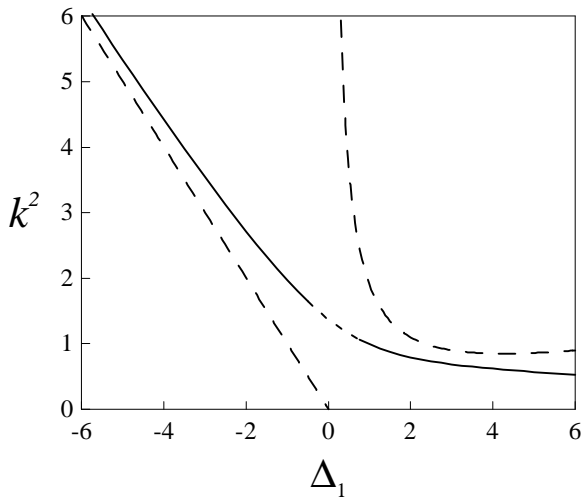

*Fig.3. The wavenumber of the patterns, given by the linear stability analysis for $a = E = 10$. The exact value is given by the solid line. Dashed lines correspond to analytical expressions given in the text.*

Some analytical expressions can be found in different limits. Assuming the pump and diffraction are large, we obtain, for negative detunings,

$$k^2 = -\Delta_1, \qquad (4)$$

that clearly corresponds to the off-resonance patterns selected by the cavity. On the other side, for positive detunings, we find

$$k^2 = \frac{\Delta_1}{a} + \frac{2E}{a\Delta_1}. \qquad (5)$$

The asymptotic expressions (4) and (5) are represented by dashed curves in Fig.3, to be compared with the exact result (full line).

Next a numerical analysis was performed, solving the DOPO equations (2) by using a split-step scheme on a spatial grid of 128X128 and periodic boundary conditions. For more details of the numerical scheme see, *e.g.*, [4]. The initial condition for the numerical integration was a randomly distributed field ($\delta$–correlated in space Gaussian noise).

In Fig.4 we evaluate the threshold for emergence of spatial patterns, for a fixed value of signal detuning. We show the results obtained from the linear stability analysis described above (full line), together with the numerical results for some values of diffraction (represented by symbols). In all cases, the final LALI patterns have hexagonal symmetry, like the one shown in the inset of Fig.4. For comparison, the preferred patterns occurring in off-resonance region are not hexagons, but stripes.

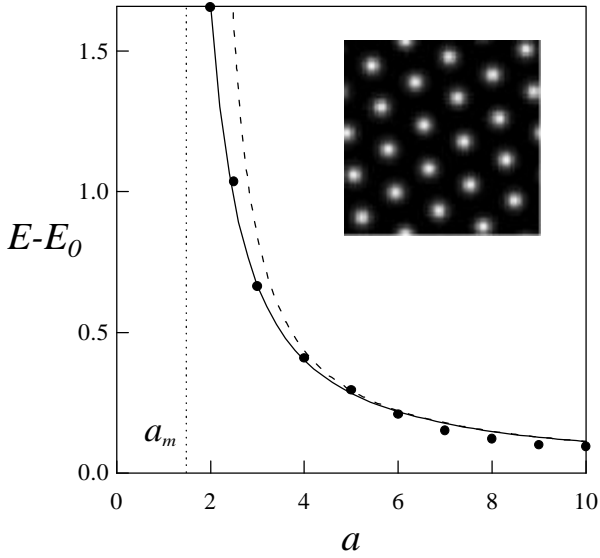

*Fig.4. Critical pump value for LALI instability depending on diffraction, for fixed signal detuning $D_1 = 2$. The symbols represent the result of the numerical integration of the DOPO equations. The solid curve is the semi-analytical calculation from the linear stability analysis, and the dashed curve corresponds to the boundary of the instability domain given by Eq.(8). The inset shows an hexagonal pattern numerically evaluated for $a = 5$, $E = 3$.*

The threshold condition for LALI instability can be evaluated analytically, but is in general a complicated function of the parameters. However, Fig.4 suggest that it follows two main features: (*i*) there exist an hyperbolic dependence between the pump and the diffraction, and (*ii*) a minimum value of the diffraction $a_m$ is required to reach the instability for fixed detuning. This guide the search for an asymptotic expression, introducing a smallness parameter related with the deviation from the threshold, given by $E_0 = \sqrt{1+\Delta_1^2}$. Then, assuming $R = E_0(E-E_0) \sim O(\varepsilon)$ and $D = a - a_m \sim O(1/\varepsilon)$, and expanding the eigenvalue we find, at leading order in $\varepsilon$, that the homogeneous solution is unstable whenever

$$27 D^2 R^2 - \Delta_1^2 (2DR - 1)^3 < 0. \qquad (6)$$

The minimum value of the diffraction $a_m$, that it depends on the detuning, can be evaluated analyzing the opposite limit, *i. e.*, at large values of the pump over threshold. Then we find that the instability can be observed only when $a > a_m$, where

$$a_m (E_0 - 1) = 1. \qquad (7)$$

From (7) it follows that $a_m$ grows monotonically when decreasing the detuning, and that $a > 1$ (and consequently $a_0 > a_1$) when $\Delta_1 < \sqrt{3}$.

Finally, the instability domain (6) can be written, in the original variables, as

$$\qquad (8)$$

where **h** is a positive function of the signal detuning. In the limit of small detuning (6) yields an asymptotic value $h = 27/8\Delta_1^2$.

The expression (8) is plotted in Fig.4 (dashed line) for $\Delta = 2$ (for which $\mathbf{h} = 2$). Notice the good correspondence with the exact (full line) and numerical (symbols) results.

Turing instability in DOPOs occurs only for nonzero signal detuning, as follows from the stability analysis and also from Eq.(6). However, in resonance, for nonzero pump diffraction some particular transverse wavenumbers are weakly damped (LALI is under-developed). The wavenumber of weakly damped modes follows from the linear stability analysis of Eqs.(2). In the limit of far above the threshold ($E \gg 1$) is given by

$$k^2 = \sqrt{\frac{2E}{a}} \qquad (9)$$

which is valid also for small values of detuning $\Delta_1$, where pattern formation is expected.

Eq.(9) means a ring of weakly damped wavevectors, in the spatial Fourier domain. To check the existence of the ring numerically one must introduce a permanent noise. One can expect that the homogeneous solution will be then weakly modulated by a filtered noise, with a characteristic wavenumber given by (9).

In order to incorporate the noise, we modify Eqs.(2), by adding the term $\sqrt{g_i}\Gamma_i$ to the evolution equation for each field component $A_i$ in Eqs.(2). These terms, introduced phenomenologically, represent stochastic Langevin forces defined by $\langle \Gamma_i(r_1,t_1) \cdot \Gamma_i^*(r_2,t_2) \rangle = \mathbf{d}(r_1 - r_2) \cdot \mathbf{d}(t_1 - t_2) / 2T_i$, where $T_i$ are the corresponding temperatures.

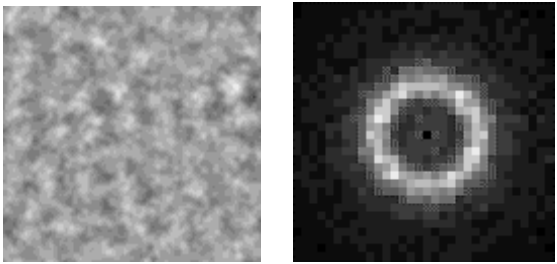

*Fig.5.* Stochastic spatial distribution (a) and averaged spatial Fourier power spectrum (b) as obtained by numerical integration of DOPO Langevin equations, for $\mathbf{D}_1 = 0$, $E = 2$, $a_0 = 0.005$, $a_1 = 0.0005$ ($a = 10$). Averaging time $t = 300$. The zero spectral component has been removed.

A typical result of numerical integration of the Langevin equations is shown in Fig.5, where a snapshot of the amplitude and the corresponding averaged spatial power spectrum $\langle |A(\mathbf{k})|^2 \rangle$ are shown. As expected, no spatial wavenumber selection was visible in the case of zero pump diffraction. For nonzero pump diffraction the DOPO filters the off-axis noise components, and a ring in the far field emerges (Fig. 5). Increasing the values of pump diffraction the induced wavenumber ring decreases in radius, and gets more dominant, in accordance with (9).

The above calculations have been performed for zero detuning for both waves. Therefore all possible pattern formation mechanisms occurring due to off-resonance excitation are excluded.

The expression (9) for the wavenumber, although evaluated at resonance, is a good approximation for the wavenumber of the patterns for moderate values of the signal detuning, and corresponds to a characteristic length of the emerging pattern $L_p = k_c^{-1}$. Returning to the initial normalizations of space in Eqs.(2), it can be expressed as

$$L_p^2 = \sqrt{\frac{a_0 a_1}{2E}}. \tag{10}$$

which, together with Eq.(8), is strikingly similar to the one derived in [8] for the Brusselator model.

Clearly, the scale of the pattern given by (10) depends on the diffraction of both fields, as a result of the interaction. It is interesting to compare the scale of the Turing pattern with the characteristic scales of the components, given by their spatial evolution in the absence of interaction. For this, we consider first a deviation with respect to the trivial solution: $A_0 = E + X$, $A_1 = Y$. In the resonant case and neglecting the nonlinear interaction, Eq.2(a) leads to

$$-Y + EY^* + ia_1 \nabla^2 Y = 0, \tag{11}$$

or, equivalently,

$$\left[1 - \left(\frac{a_1}{E}\nabla^2\right)^2\right]Y = 0. \tag{12}$$

Similarly, from Eq.2(b) we find

$$-X + ia_0 \nabla^2 X = 0. \tag{13}$$

From the solutions of (12) and (13) we can define a characteristic spatial scale for the activator, $L_a = \sqrt{a_1/E}$, and for the inhibitor, $L_i = \sqrt{a_0}$, corresponding to the signal and pump fields respectively. Now, the scale of the generated pattern can be written in terms of the scales of the activator and the inhibitor, as

$$L_p^2 = \frac{L_a L_i}{\sqrt{2}}, \tag{14}$$

revealing that the characteristic spatial scale of the pattern is the geometric mean of the spatial scales of the interacting components.

It is possible to find a simple relation between $L_a$ and $L_i$ in the limit in which $a \gg a_m$ and $E \gg E_0$ (large diffraction and pump, and moderate detuning). Then, the instability domain (8) obtains the form $aE > h$, which can be expressed in terms of the characteristic lengths to give the threshold condition

$$L_i > \sqrt{h} L_a. \tag{15}$$

The value of $h$ depends on signal detuning and can be easily evaluated from (6). We find that $\sqrt{h} > \sqrt{2}/2$ always, and in particular, that $h > 1$ for $\Delta_1 < 3\sqrt{3}$. Then, for small (and also moderate) detunings the inhibitor range must be larger than the activator range for occurrence of LALI instability. This is in accordance with the assumptions made in the derivation of (9).

The conditions defined by Eqs.(14) and (15) are typically found in reaction-diffusion systems, and are a signature of the Turing character of the instabilities described.

Concluding, we have shown that the interplay between diffractions of interacting field components can cause a pattern-forming instability in DOPOs. The reported mechanism is analogous to that of local activation and lateral inhibition first described by Turing [6]. The difference between the original LALI instability of Turing and LALI instability reported here for DOPOs is that in the latter case it is related to the interplay between diffractions, and not diffusions of interacting components. However, the analogy is not complete. In optics, the fields (playing the role of concentrations) are complex, and consequently the number of species is larger. Also, the presence of detunings result in complex coefficients (which could be compared with the kinetic constants). We expect then a richer variety of phenomena, together with some quantitative deviations. For example, Turing patterns in optics are possible even when $a_0 < a_1$, which is not possible in the chemical analogue. In fact, hexagonal patterns for $a = 1/2$ and positive detuning were predicted before [18-20], although their origin was not related to the Turing mechanism described in this paper.

We think that this mechanism is rather general to nonlinear optics, and should be present in systems with other nonlinearities, different from the second-order one shown here. The analysis of other models is under current progress.

Both diffraction and diffusion are different representations of nonlocalities of interacting components. Therefore we generalize the Turing pattern formation mechanism: not necessarily the interplay between different diffusions, but in general the interplay between different nonlocalities, can induce pattern formation.

We acknowledge discussions with C.O.Weiss, G.J. de Valcárcel and E. Roldán. This work has been supported by Acciones Integradas Project HA1997-0130, NATO grant HTECH.LG 970522, Sonderforschungs Bereich 407 and by DGICYT of the Spanish Government under grant nº. PB98-0935-C03-02.

References

1. P. Mandel, M. Georgiou, and T. Erneux, Phys. Rev. A **47**, 4277 (1993).
2. W.J. Firth and A.J. Scroggie, Europhys. Lett. **26**, 521 (1994).
3. K. Staliunas, M.F.H. Tarroja, G. Slekys, C.O. Weiss, and L. Dambly, Phys. Rev. A **51**, 4140 (1995).
4. G.J.de Valcarcel, K.Staliunas, E.Roldan, and V.J.Sanchez-Morcillo, Phys.Rev. A **54**, 1609 (1996).
5. S. Longhi, Phys. Rev. A. **53**, 4488 (1996).
6. L. A. Lugiato and R. Lefever, Phys. Rev. Lett. **58**, 2209 (1987).
7. A.M. Turing, Phil. Trans. Roy. Soc. Lond. B **237**, 37 (1952).

8. I. Prigogine and R. Lefever, J. Chem. Phys. **48**, 1696 (1968).
9. H. Meinhardt, Models of biological pattern formation (Academic, London, 1982).
10. J.D. Murray, Mathematical Biology (Springer, Berlin, 1989).
11. M.C. Cross and P.C. Hohenberg, Rev. Mod. Phys. **65**, 851(1993).
12. G.L. Oppo, M. Brambilla and L.A. Lugiato, Phys. Rev. A. **49**, 2028 (1994).
13. V.B. Taranenko, K. Staliunas and C.O. Weiss, Phys. Rev. A. **56**, 1582 (1997); C.O. Weiss, M. Vaupel, K. Staliunas, G. Slekys and V.B. Taranenko, Appl. Phys. B **68**, 151 (1999).
14. V.B. Taranenko, K. Staliunas and C.O. Weiss, Phys. Rev. Lett. **81**, 2236 (1998).
15. L.A. Lugiato, C. Oldano, C. Fabre, E. Giacobino, and R. Horowicz, Nuovo Cimento **10**D, 959 (1988).
16. K. Staliunas, and V.J. Sanchez-Morcillo, Phys. Rev. A. **57**, 1454 (1998).
17. S. Longhi, Physica Scripta, **56**, 611 (1997).
18. M. Brambilla, D. Camesasca, and G.L. Oppo (unpublished).
19. K. Staliunas, and V.J. Sanchez-Morcillo, Opt. Commun. **139**, 306 (1997).
20. M. Le Berre, D. Leduc, E. Ressayre and A. Tallet, J. Opt. B: Quantum Semiclass. Opt. **1**, 153 (1999).